\documentclass[final]{aipproc}
\layoutstyle{6x9}
\usepackage{multirow}

\newcommand{\drbar}{{\overline{\rm DR}}}
\newcommand{\mhs}{{MH}}
\newcommand{\atts}{{\mathrm{A}_{\tau \tau}}}

\newcommand{\neuto}{{\tilde{\chi}}_1^0}

\newcommand{\Omegah}{\Omega_{\chi} h^{2}}

\begin{document}

\title{Coannihilation with a chargino and gauge boson pair production at one-loop}
\rightline{PITHA 09/22}
\rightline{LAPTH-Conf-1352/09}

\classification{95.35.+d, 98.80.-k, 11.30.Pb, 11.10.Gh}

\keywords      {dark matter, supersymmetry, loop calculation, cosmology}

\author{\underline{N.~Baro}}{
  address={Institut f\"{u}r Theoretische Physik E, RWTH Aachen University, D-52074 Aachen, Germany}
}

\author{G.~Chalons}{
address={LAPTH, Universit\'e de Savoie, CNRS, BP 110, F-74941 Annecy-le-Vieux Cedex, France}
}

\author{Sun Hao}{
address={LAPTH, Universit\'e de Savoie, CNRS, BP 110, F-74941 Annecy-le-Vieux Cedex, France}
}

\begin{abstract}
We present a complete calculation of the electroweak one-loop corrections to the relic density within the MSSM framework. In the context of the neutralino as dark matter candidate, we review different scenarios of annihilation and coannihilation with a chargino. In particular we investigate predictions for the annihilation into gauge boson pairs for different kinds of neutralino: bino-, wino- and higgsino-like.
We present some interesting effects which are not present at tree-level and show up at one-loop. To deal with the large number of diagrams occuring in the calculations, we have developed an automatic tool for the computation at one-loop of any process in the MSSM. We have implemented a complete on-shell gauge invariant renormalization scheme, with the possibility of switching to other schemes. We emphasize the variations due to the choice of the renormalization scheme, in particular the one-loop definition of the parameter $\tan\beta$.
\end{abstract}

\maketitle

The elucidation of the nature of the Dark Matter is one of the most challenging issues in the incoming years. The relic density of Dark Matter is now well known thanks to the WMAP and other experiments, $0.098 < \Omega h^2 < 0.122$, and we expect that the PLANCK satellite will measure this density with a precision of only $2\%$. As concerns cosmology this shows that we are entering an era of precision measurements. The phenomenological predictions in New Physics should reach the same level of precision inducing the need to go at least at the one-loop level. The Minimal Supersymmetric Standard Model (MSSM) is a well-established and coherent theory which explains many of the problems of the Standard Model beside providing a possible candidate for Dark Matter, the lightest neutralino $\tilde{\chi}_{1}^{0}$.

To explain the excess reported by the PAMELA collaboration\cite{pamela}, it was argued this could be interpreted as a signal from a wino-like Light Supersymmetric Particle (LSP)\cite{kane08}. So the MSSM parameter space has to fulfill $M_{2} < M_{1}, |\mu|$ and consequently the mass of the lightest neutralino is approximately the same as the lightest chargino mass which is given mainly by the $M_{2}$ parameter. The chargino $\tilde{\chi}_{1}^{+}$ is the Next-to-LSP and therefore the neutralino-chargino coannihilation could play an important role for the relic density. In the following we will not restrict ourselves to a wino-like neutralino but consider different kinds of neutralino where coannihilation is important. As mentioned before a high precision is required in relic density and this study is then devoted to the one-loop corrections to the cross-sections that enter the annihilation into gauge bosons. 

An automatic tool, {\tt SloopS}, is used to deal with the large number of diagrams involved in such calculations. The renormalization scheme and different tests have been published in Ref.~\cite{sloops} and first results concerning Dark Matter at one-loop have been reported in Ref.~\cite{baro07}. Different checks are done systematically on our results such as Ultra-Violet and Infra-Red finiteness, cut-off independence and also gauge invariance. In {\tt SloopS} we have implemented a complete renormalisation of the MSSM in all the sectors with an On Shell scheme as a default. As a bonus we can investigate various renormalization schemes in particular for the important and ubiquitous $\tan\beta$ which represents at tree-level the ratio of the vacuum expectation values of the two Higgs doublets\footnote{\protect\label{footschemes}At the one-loop level we choose $3$ definitions: $\atts$ where the parameter is extracted from the decay of the $CP$-even Higgs $A^{0}\rightarrow \tau^+ \tau^-$, $\drbar$ where $\tan\beta$ is a pure divergence at one-loop and $\mhs$ where the heaviest neutral Higgs $H^0$ is taken as an input parameter\cite{sloops}.}. 

\begin{figure}
\begin{tabular}{lcccccc} 
&\footnotesize $M_1$ &\footnotesize $M_2$ &\footnotesize $\mu$ &\footnotesize $M_{A^0}$ &\footnotesize $\tan\beta$  &  \multirow{12}{*}{\includegraphics[height=.28\textheight]{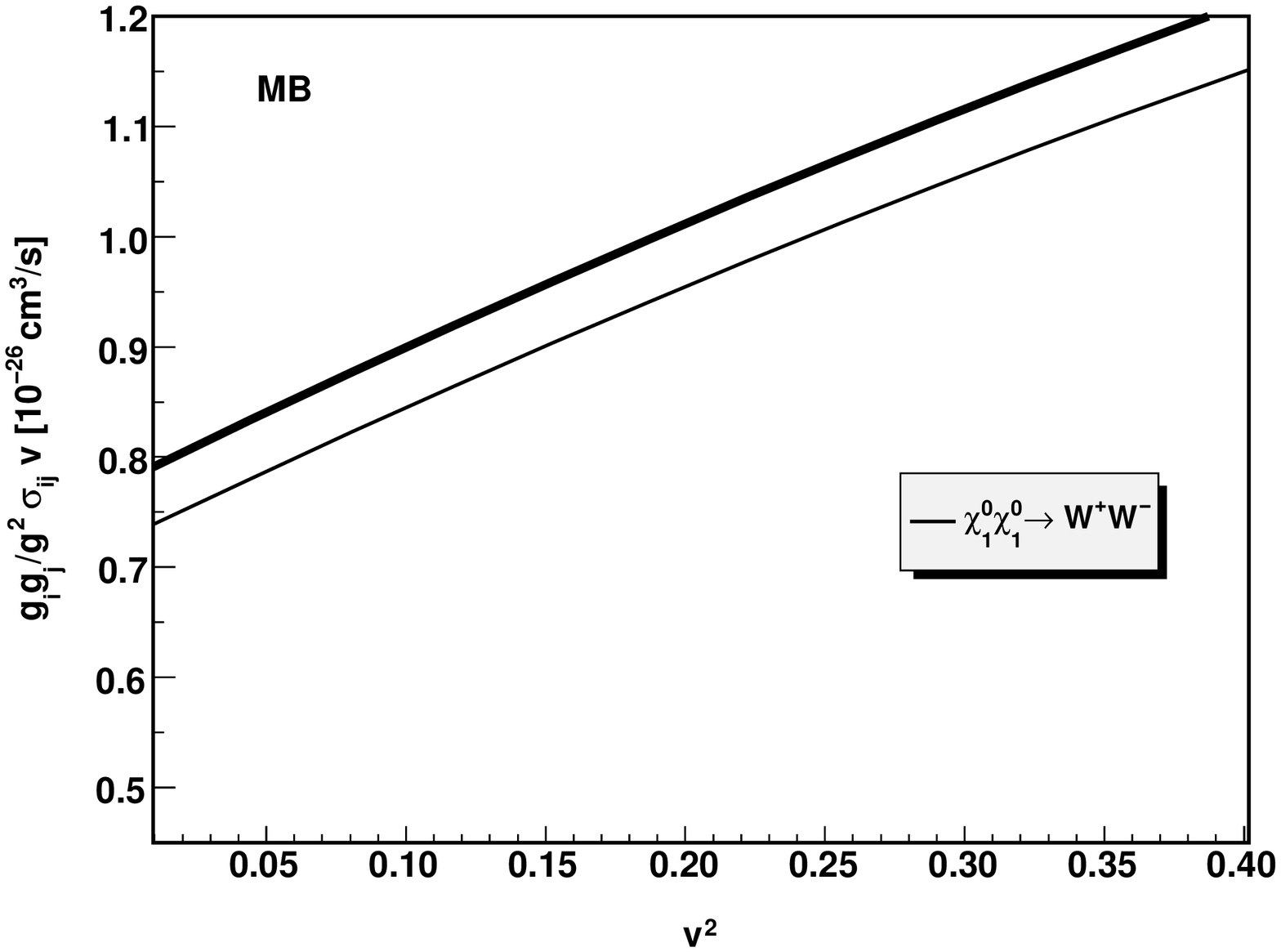}}
\\
\cline{1-6}
\footnotesize Mixed-Bino [MB] &\footnotesize 110  &\footnotesize 134.5 &\footnotesize -245 &\footnotesize 600 &\footnotesize 10  & \\
\footnotesize Higgsino [H]  &\footnotesize 400  &\footnotesize 350   &\footnotesize -250 &\footnotesize 800 &\footnotesize 4  & \\
\footnotesize Mixed-Higgsino [MH] &\footnotesize 565  &\footnotesize 1000  &\footnotesize 550  &\footnotesize 1350 &\footnotesize 4 & \\
\footnotesize Light-Wino [LW] &\footnotesize 550  &\footnotesize 210   &\footnotesize -600 &\footnotesize 700 &\footnotesize 30  & \\
\footnotesize Heavy-Wino [HW] &\footnotesize 3500 &\footnotesize 1800  &\footnotesize 4500 &\footnotesize 5000 &\footnotesize 15 & \\
\cline{1-6}
&\footnotesize $M_3$ &\footnotesize $M_{\tilde{q}_L}$ &\footnotesize $M_{\tilde{l}_L}$ &\footnotesize $M_{\tilde{f}_R}$ &\footnotesize $A_f$ & \\
\cline{1-6}
\footnotesize Mixed-Bino [MB] &\footnotesize 600  &\footnotesize 600  &\footnotesize 600  &\footnotesize 600  &\footnotesize 0  & \\
\footnotesize Higgsino [H]  &\footnotesize 1000 &\footnotesize 650  &\footnotesize 650  &\footnotesize 650  &\footnotesize 0  & \\
\footnotesize Mixed-Higgsino [MH] &\footnotesize 1200 &\footnotesize 1700 &\footnotesize 1700 &\footnotesize 1700 &\footnotesize 0  & \\
\footnotesize Light-Wino [LW] &\footnotesize 1200 &\footnotesize 387  &\footnotesize 360  &\footnotesize 800  &\footnotesize 0  & \\
\footnotesize Heavy-Wino [HW] &\footnotesize 5000 &\footnotesize 5000 &\footnotesize 5000 &\footnotesize 5000 &\footnotesize 0  & \\
\cline{1-6} 
\multicolumn{6}{c}{\includegraphics[height=.28\textheight]{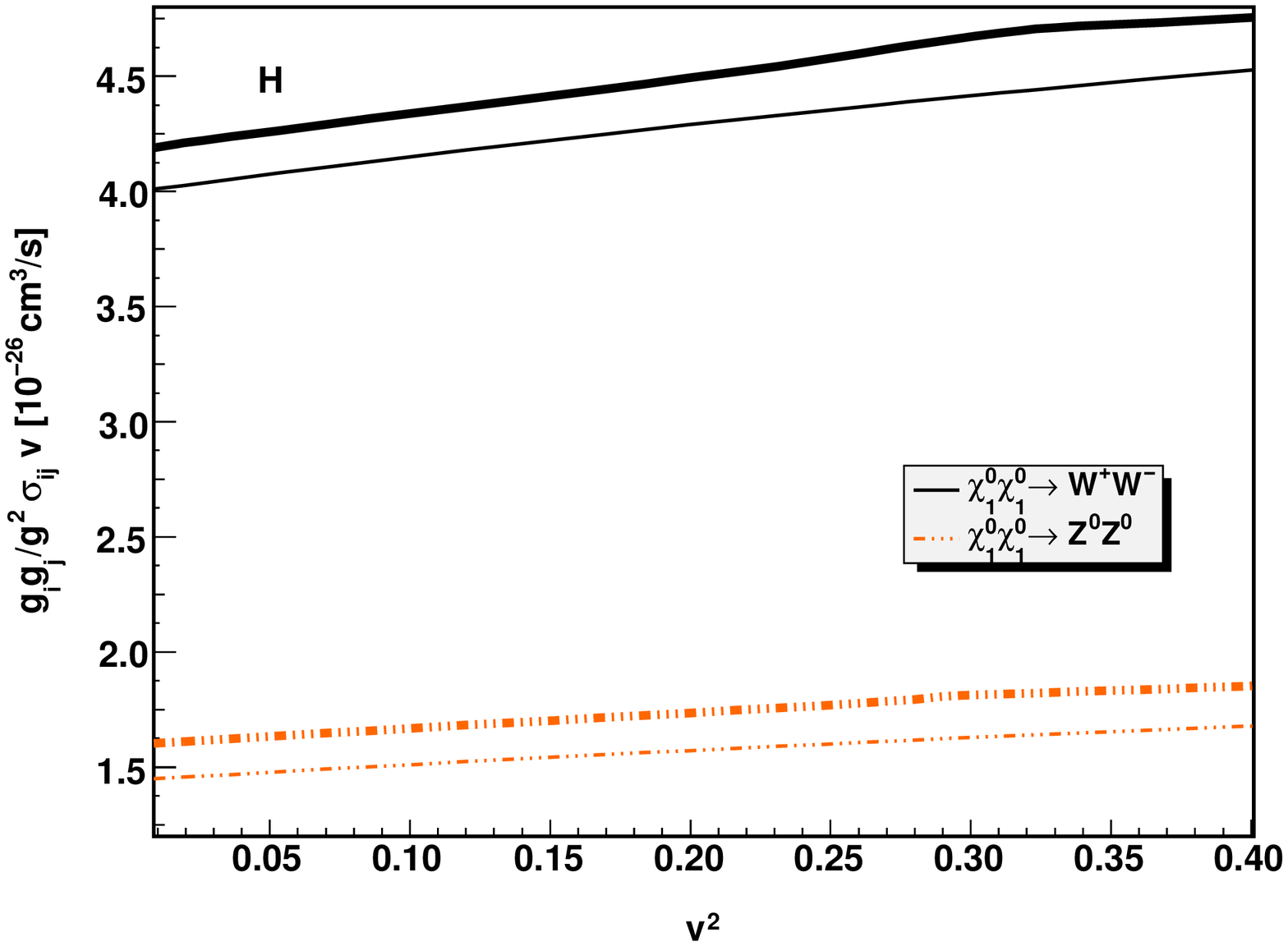}} & \includegraphics[height=.28\textheight]{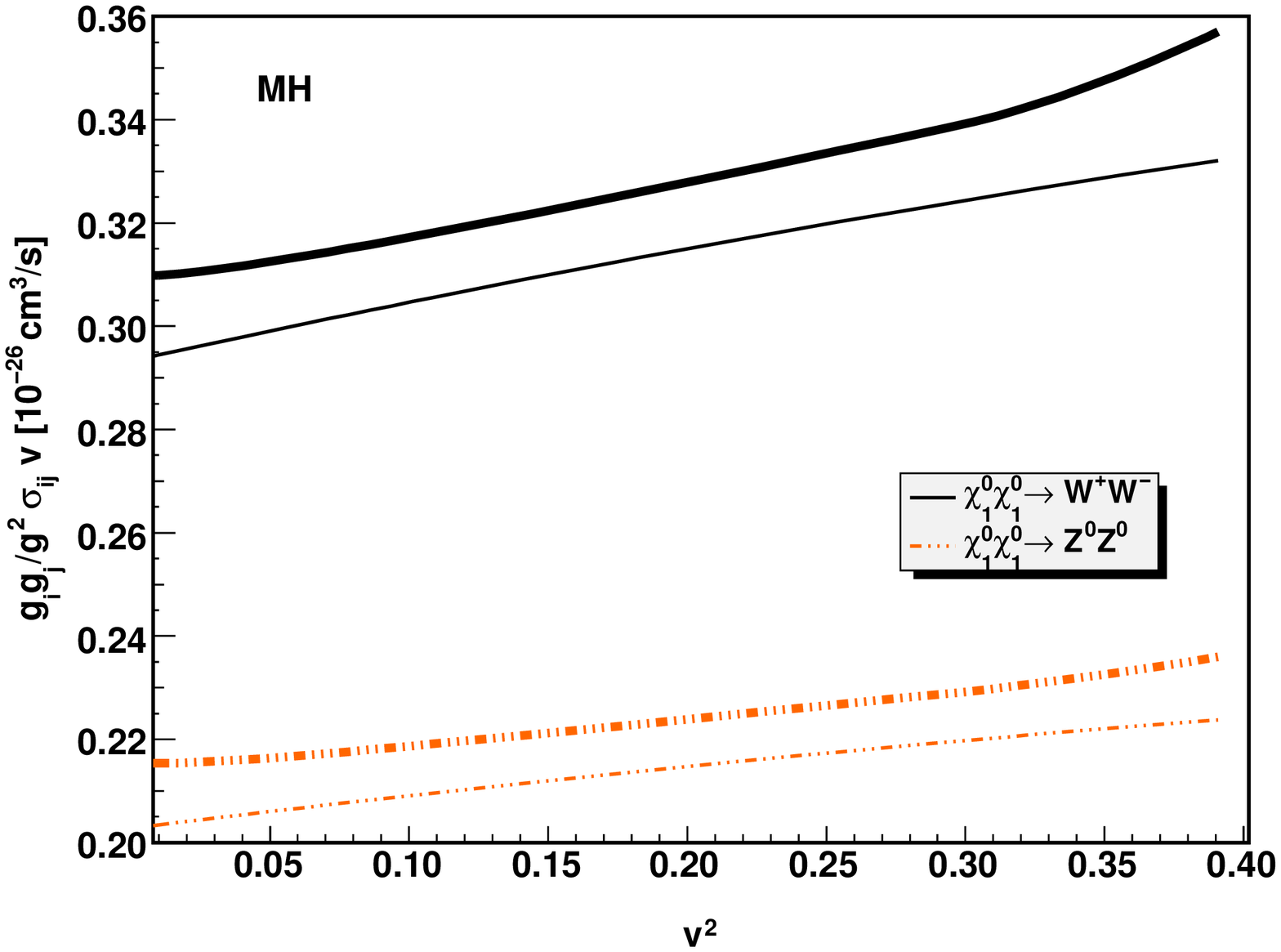} \\
\multicolumn{6}{c}{\includegraphics[height=.28\textheight]{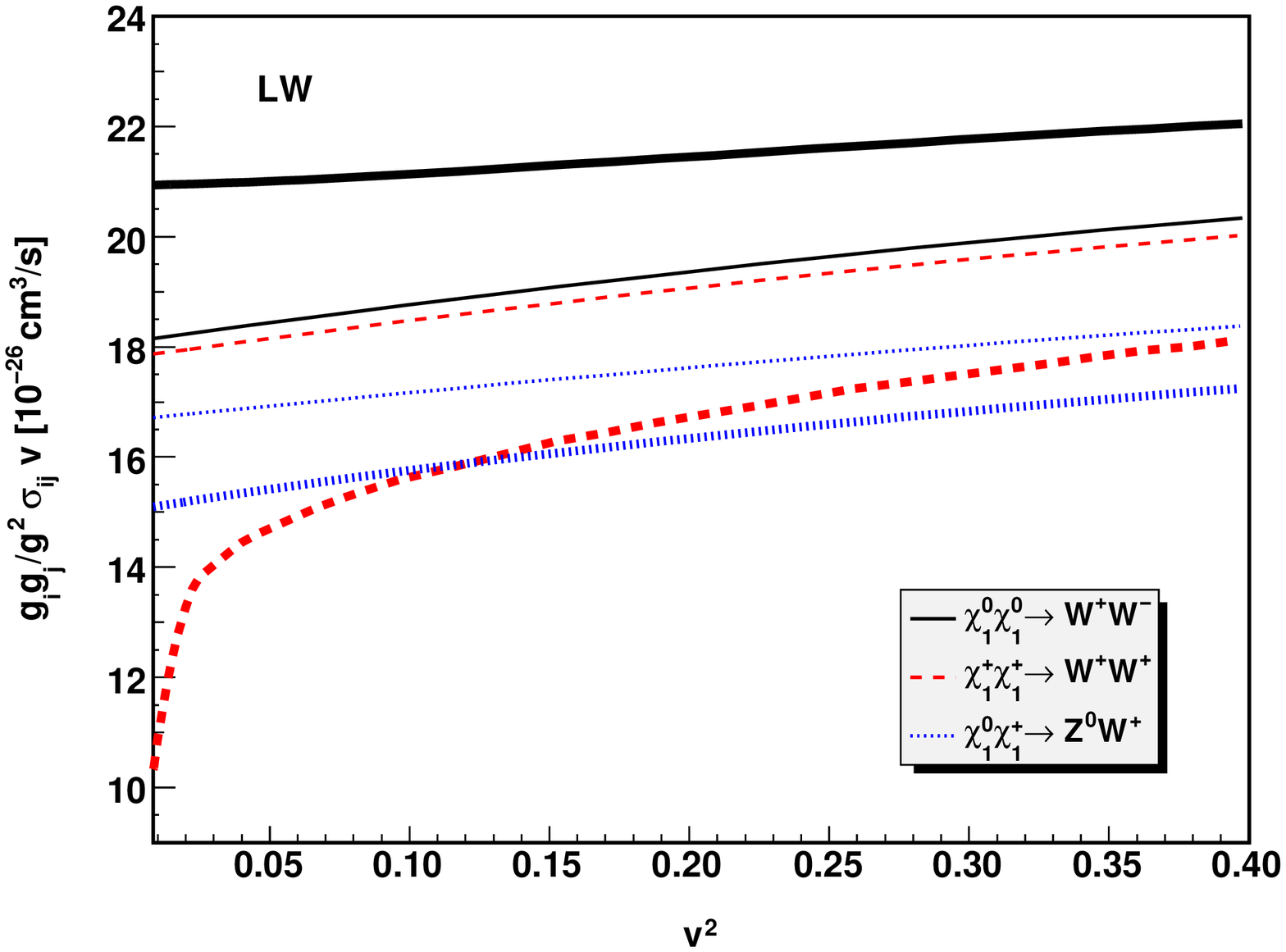}} &   \includegraphics[height=.28\textheight]{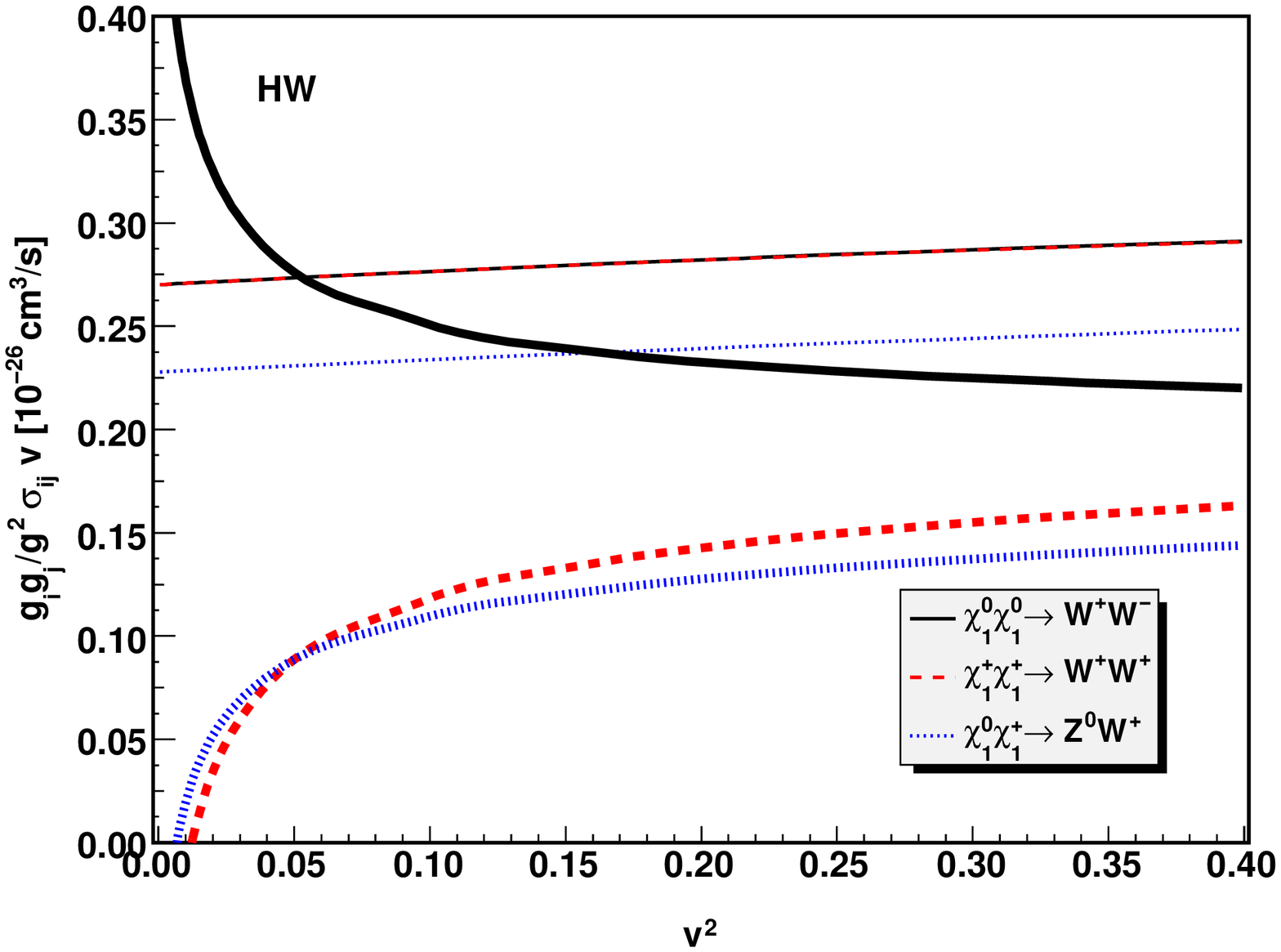}
\end{tabular}
\caption{Main contributions to the relic density in various scenarios: Mixed-Bino [MB], Higgsino [H], Mixed-Higgsino [MH], Light-Wino [LW] and Heavy-Wino [HW]. 
The upper left panel shows our sets of parameters. Masses are in GeV.
For each process the thin curve represents the tree-level cross-section whereas the thick curve shows the one-loop level one in an On-Shell scheme ($\atts$). 
To emphasize the connexion to the relic density calculation, the cross-sections are normalized by a factor which takes into account all the effective degrees of freedom. This makes clear the relative contributions of these processes to the relic density of Dark Matter. Explicitly the $g_{i}^{\textrm{\tiny eff}}g_{j}^{\textrm{\tiny eff}}/g^{\textrm{\tiny eff}\; 2}$ factor is such as $g^{\textrm{\tiny eff}} = \sum_{i} g_{i}^{\textrm{\tiny eff}}$ and $g_{i}^{\textrm{\tiny eff}}$ is given by
$
 g_{i}^{\textrm{\tiny eff}} = \frac{g_{i}}{g_{0}} \left( 1 + \Delta m_{i} \right)^{3/2} \exp(-x \Delta m_{i})$, $\Delta m_{i} = (m_{i}-m_{\tilde{\chi}_{1}^{0}})/m_{\tilde{\chi}_{1}^{0}}$. $x=m/T$ is taken at the freeze-out temperature.
 } \label{fig:all}
\end{figure}

In order to turn the cross-sections into a precise prediction for the relic density of Dark Matter, we interfaced {\tt SloopS} with {\tt MicrOmegas} which handles the cosmological part of the calculation (we plug in all the channels accounting for at least $5\%$ of the total amount of the relic density and not just the dominant vector boson channels). We also rewrite a tree-level model in {\tt MicrOmegas} to obtain the relic density without any improvements such as radiative mass corrections or effective couplings.
We present here some scenarios covering different aspects at one-loop of the neutralino-chargino coannihilation in the MSSM for different decomposition of the lightest neutralino: mixed-bino like neutralino [MB], higgsino-like neutralino [H], mixed-higgsino like [MH], light wino like [LW] and heavy-wino like neutralino [HW], see Fig.~\ref{fig:all}. The Standard Model parameters are the same as in Ref.~\cite{baro07}. Figure \ref{fig:all} summarizes our findings for the gauge boson pair production at tree-level as well as at the one-loop level in the $\atts$ schem$\textrm{e}^{\protect\ref{footschemes}}$.

\noindent
\textbf{Mixed-Bino:} The chargino coannihilation is completely suppressed in a pure bino scenario because of the large chargino mass. To make it important for a bino neutralino one needs to have a wino or higgsino mixed neutralino by taking for example $M_{1} \sim M_{2}$. The contribution to the relic density is then as expected into gauge bosons and light quarks
\footnote{We also consider EW/QCD one-loop corrections to the channel $\tilde{\chi}_{1}^{0}\tilde{\chi}_{1}^{+}\rightarrow u\bar{d} / s\bar{c}$, see Ref.~\cite{baro09}.}. The main channel $\tilde{\chi}_{1}^{0}\tilde{\chi}_{1}^{0}\rightarrow W^{+}W^{-}$ accounts at tree-level for $44\%$ of the total relic density. The product $\sigma v$ is well approximated by the expansion in terms of the square of the relative velocity $a+b v^2$ where $a$ gives the contribution of the $s$-wave and $b$ of the $p$-wave. In Tab.~\ref{tab:mixedbino} we summarize the correction to $a$ and $b$ for the different $\tan \beta$ renormalization schemes. One can observe a strong dependence on the renormalization of $\tan\beta$. 
\begin{table}[h!]
\begin{tabular}{lcccccc}
\hline
& & Tree-Level & & $\mathrm{A}_{\tau \tau}$ & $\drbar$ & MH \\
\hline
\multicolumn{1}{l}{$\neuto\neuto\rightarrow W^{+} W^{-}$ $[44\%]$ }
                & $a$ &  $+0.810$  &   & $+7.60\%$  &$+12.16\%$& $+29.60\%$ \\
                & $b$ &  $+1.219$  &   & $+0.78\%$ & $+7.10\%$ & $+24.20\%$ \\
\hline
 $\Omegah$      &    &  $0.108$    &   & $0.105$ ($-2.8\%$)  & $0.102$ ($-5.6\%$) & $0.097$ ($-10.2\%$) \\
\hline
\end{tabular}
\caption{
{\em Mixed-Bino scenario: Tree-level values of the $a$ and $b$ coefficients in units $ 10^{-26} {\rm cm}^{3}
{\rm s}^{-1}$, as well as the relative one-loop corrections in the
$\mathrm{A}_{\tau \tau}$, $\drbar$, and $\mhs$ scheme.}}\label{tab:mixedbino}
\end{table} 

\noindent
\textbf{Higgsino:} In this scenario $|\mu| < M_{1},M_{2}$, the neutralino is now mostly higgsino and as the $\tilde{\chi}_{1}^{0}\tilde{\chi}_{2}^{0}Z^{0}$ and $\tilde{\chi}_{1}^{0}\tilde{\chi}_{1}^{\pm}W^{\mp}$ couplings are large, the main annihilation channels are $\tilde{\chi}_{1}^{0}\tilde{\chi}_{1}^{0}\rightarrow W^{+}W^{-} [26\%] / Z^{0}Z^{0} [9\%]$. The $\tan\beta$-scheme dependence is of about $1\%$ at most. The value of the standard thermal relic density is $\Omega h^{2} = 0.00931$ and receives a correction of about $-2.4\%$ in the $\atts$ scheme. This density is of course too small but in other cosmological scenarios as non-thermal relic, the corrections we get remain interesting and should be taken into account. 

\noindent
\textbf{Mixed-Higgsino:} As a pure higgsino-like neutralino annihilates too efficiently into gauge bosons we reduce the higgsino fraction by taking a bino-higgsino-mixed neutralino, $|\mu| \sim M_{1}$. The main channels are the annihilation into $W^{+}W^{-}$ $[19\%]$ and $Z^{0}Z^{0}$ $[13\%]$ final states. We also observed a weak $\delta \tan\beta$ dependence. The relic density is $\Omega h^{2} = 0.0814$ with a correction of $- 1.2\%$.

\noindent
\textbf{Light-Wino:} A wino neutralino is obtained by taking $M_{2} < M_{1}, |\mu|$. At tree-level, the two processes $\tilde{\chi}_{1}^{0}\tilde{\chi}_{1}^{0}\rightarrow W^{+}W^{-} [13\%]$ and $\tilde{\chi}_{1}^{+}\tilde{\chi}_{1}^{+}\rightarrow W^{+}W^{+} [12\%]$ are almost degenerate. This is due to the fact that in a wino scenario only the $\tilde{\chi}_{1}^{0}\tilde{\chi}_{1}^{+}W^{+}$ coupling is relevant. The main contribution comes from the exchange of a chargino/neutralino via the t-channel. The radiative corrections lift this degeneracy at the one-loop level. In addition an interesting singularity shows up for $\tilde{\chi}_{1}^{+}\tilde{\chi}_{1}^{+}$. This correction is the one-loop manifestation of the non-relativistic Coulomb-Sommerfeld effect due to the exchange of a photon between the two incoming charged particle. The thermal relic is $\Omega h^{2} = 0.00215$ with a correction of $- 1.9\%$ with no $\tan\beta$-scheme dependence.

\noindent
\textbf{Heavy-Wino:} This scenario is almost the same as the latter but here we tried to have an allowed relic density at the born level with a neutralino mass of $1800$ GeV and a heavy spectrum. All the cross-sections get now a singularity at one-loop when the velocity becomes small. The linear behaviour $\sigma v = a+b v^2$ is no longer valid for such a scenario with a heavy wino neutralino and loop corrections are large for small velocity. The singularity is the non-perturbative effect from the Coulomb-Sommerfeld interaction coming mainly from the exchange of the gauge particles $W^{\pm}$ and $Z^0$ between the two initial particles. The relic density gets a sizeable correction $\Omega h^{2} = 0.0997 + 9.3\%$.\\

We would like to thank Fawzi Boudjema for a careful reading of the manuscript. This work is part of the French ANR project, {\tt ToolsDMColl} BLAN07-2 194882 and is supported in part by GDRI-ACPP of the CNRS (France). This work is also supported in part by the European Community's Marie-Curie Research Training Network under contract MRTN-CT-2006-035505 ``Tools and Precision Calculations for Physics Discoveries at Colliders'', the DFG SFB/TR9 ``Computational Particle Physics'', and the Helmholtz Alliance ``Physics at the Terascale''.

\bibliographystyle{aipproc}

\end{document}